\documentclass[aps,twocolumn,showpacs]{revtex4}

\usepackage{graphicx}

\begin{document}

\title{Compact-like discrete breathers in systems with \\ nonlinear and
nonlocal dispersive terms}

\author{A.V. Gorbach}
\affiliation{Max-Planck-Institut f\"ur Physik komplexer Systeme,
N\"othnitzerstr. 38, Dresden 01187, Germany}

\author{S. Flach}
\affiliation{Max-Planck-Institut f\"ur Physik komplexer Systeme,
N\"othnitzerstr. 38, Dresden 01187, Germany}

\begin{abstract}
Discrete breathers with purely anharmonic short-range interaction potentials
localize {\it super-exponentially} becoming compact-like.
We analyze their spatial localization properties and their
dynamical stability. Several branches of solutions are identified.
One of them connects to the well-known Page and Sievers-Takeno lattice
modes, another one connects with the compacton solutions of Rosenau.
The absence of linear dispersion allows for extremely long-lived
time-quasiperiodic localized excitations. 
Adding long-range anharmonic interactions leads to an extreme case
of competition between length scales defining the spatial breather localization.
We show that short- and long-range interaction terms 
competition results in the appearance of several characteristic cross-over lengths 
and essentially breaks the concept of {\it compactness} of the 
corresponding discrete breathers.
\end{abstract}

\pacs{63.20.Pw, 63.20.Ry, 05.45.-a}


\maketitle

\section{Introduction}

Energy localization due to a nonlinearity in dynamical systems 
has been observed for 
more than one century \cite{Russel}, and the effect of an exact balance 
between nonlinearity and linear
dispersion of wave packets leading to the appearance of soliton excitations has become a 
paradigmatic example in nonlinear science which can be found in various textbooks.
In the past decade remarkable achievements in the study of localized nonlinear excitations
were made with the discovery of stable localized modes in spatially discrete
translationally invariant Hamiltonian systems - {\it discrete breathers} (DBs) \cite{DB_rev}.
They have been proved to be generic {\it exact} time-periodic solutions of the corresponding
coupled nonlinear ordinary differential equations, eventhough the latter are generally
non-integrable. It is worth mentioning that discrete breathers
have been observed experimentally in various physical systems including
coupled optical waveguide arrays \cite{ESM+98,SKE+03,FSE+03}, coupled Josephson junctions \cite{TMO00},
micromechanical cantilever systems \cite{SHS+03,SHE+03,SHS+04},
anti-ferromagnetic crystals \cite{SES99,SS04}, high-$T_c$ superconductors
\cite{MK04}. Discrete breathers are predicted also
to exist in the dynamics of dusty plasma crystals \cite{KS05}.

Among the most important characteristics of a localized excitation are its
localization length and the spatial decay characteristics of its amplitude.
Although DBs can be localized practically on a single site, 
in most of the cases they have exponentially decaying tails (similar to their 
continuum counterparts - solitons).
This is true if the interaction potentials are reasonably short ranged (see
\cite{Fla98} for details).
However, when an anharmonic interaction 
between adjacent sites is much stronger then the harmonic one, localized excitations 
can become even more compact. As it was demonstrated by Rosenau and Hyman
\cite{RH93, Ros94}, in continuous systems nonlinear localized excitations may
compactify, i.e. gain strictly zero tails, under nonlinear
dispersion. The same was conjectured for discrete
systems \cite{Kiv93}, however later it was shown 
that in discrete
systems localized excitations cannot have an exact compact structure
\cite{Fla94}, but the tail decay follows a \emph{super-exponentional}
law $e^{-a\exp bn}$, provided that the interaction is purely
short-range. This fact was then confirmed numerically
\cite{DEF+02}, and the corresponding breather solutions were coined
\emph{compact-like} \cite{DEF+02} or \emph{almost-compact} \cite{RS05} DBs. 

If compact-like breathers are dynamically stable, we may expect
that localized perturbations of such solutions will lead
to a quasiperiodic in time evolution, which will not induce
a radiation of energy away from the breather. This is in contrast
to the well-known existence of such a radiation 
for systems with linear dispersion \cite{fw93fwo94}. There it appears due to
the resonant overlap of combination frequencies of the internal perturbed
breather dynamics with the spectrum of small amplitude plane waves.
In the case of purely nonlinear dispersion the width of this spectrum
is zero, and thus the origin of the radiation is removed.
One expects then that perturbed compact-like breathers will
not radiate energy away, giving rise to genuine \emph{quasiperiodic}
compact-like breathers. 

Another important issue which might drastically change the rate of spatial decay in DBs tails
is the presence of long-range interactions,
essential e.g. in systems with weakly
screened Coulomb interaction such as ionic crystals, or
various biomolecules. Usually decaying slower than exponentially in space, 
long-range interactions
introduce a crossover length as a result of competition of the two 
essentially different length scales
\cite{Fla98}. They can also lead to the appearance of energy
thresholds for DBs in some cases, where a pure short-range 
interaction would not be capable of producing any. In
\cite{JGC+98} it was demonstrated, that the effect of
length-scale competition  with long-range algebraically and
exponentially decaying interaction can lead to a new type 
of multistability of DBs,
when in a certain model parameter regime several different 
types of DBs coexist having the same value of the spectral 
parameter (i.e. velocity or frequency).

It is the purpose of this paper to address the abovelisted issues.
The paper is organized as follows. In Section~\ref{sec:two} we introduce the
model. We
demonstrate, that the specifically chosen nonlinear potentials allow one to
completely separate temporal and spatial dependencies and thus significantly simplify
the analysis of the problem. We derive the nonlinear coupled algebraic equations for the
spatial profile of a solution and in addition an ordinary differential 
equation (Duffing equation) for the master function describing uniform
oscillations of all the sites with time. In 
Section~\ref{sec:three} we obtain the different types of discrete
breather solutions. We demonstrate, that in
general the model supports two classes of discrete breathers with completely
different dynamical properties. We then study the linear stability
properties of basic types of DBs and observe quasiperiodic localized
excitations. In Section~\ref{sec:four} we reveal the
effect of long-range interactions along the chain on properties of DB
solutions. We show that the presence of non-local dispersive terms result in the
appearance of several characteristic cross-over lengths. We derive estimations
for these cross-over lengths, as well as asymptotes for amplitude distribution
in DB tails, on the basis of a simple three-site model. In
Section~\ref{sec:concl} we conclude.

\section{The Model}
\label{sec:two}

We consider a simple
one-dimensional model of (nonlinearly) coupled oscillators
with the following Hamiltonian:
\begin{eqnarray}
\label{hamilt}
H&=&\sum_n h_n\equiv\sum_n\left\{\frac{\dot{u}_n^2}{2}+V(u_n)\right.\\
\nonumber
& &+\left.
\sum_{l>0}\frac{K}{4 l^s}\left[
(u_{n+l}-u_n)^4+(u_{n-l}-u_n)^4\right]\right\}
\end{eqnarray}
where $u_n(t)$ is the displacement of $n$th unit mass oscillator
from its equilibrium position, the constant $s>0$ characterizes
the rate of spatial decay of long-range interactions between
oscillators, and $V(u_n)$ is given by
\begin{equation}
\label{onsite}
V(u_n)=\frac{u_n^2}{2}-\frac{u_n^4}{4}.
\end{equation}
The equation of motion for the displacement of the $n$th oscillator 
from its equilibrium reads:
\begin{eqnarray}
\label{mot}
\ddot{u}_n&=&K\sum_{l}\frac{1}{l^s}\left\{\left(u_{n+l}-u_n\right)^3+
\left(u_{n-l}-u_n\right)^3\right\}\\
\nonumber
& &-u_n + u_n^3.
\end{eqnarray}
We note that while the interaction decays algebraically for any finite
power $s$, in the limit $s \rightarrow \infty$ we recover the case
of short-range nearest neighbor interaction.

The specifically chosen nonlinear potentials allow one
to use the time-space separation technique \cite{Kiv93, Fla94},
so that time-periodic solutions of (\ref{mot}) can be written in the form
\begin{equation}
\label{time-space}
u_n(t)=\phi_n G(t),
\end{equation}
with time-independent amplitudes $\phi_n$ and a master 
function $G(t)$ describing \emph{uniform oscillations} 
of all the sites. After substitution of the ansatz (\ref{time-space}) 
into the Eqs.~(\ref{mot}) the following equation for the
function $G(t)$ is obtained:
\begin{equation}
\label{ODE}
\ddot{G}+G=-CG^3,
\end{equation}
while the amplitudes $\phi_n$ satisfy algebraic equations:
\begin{eqnarray}
\label{algebra}
C\phi_n&=&-K\sum_{l>0}\frac{1}{l^s}\left[(\phi_{n+l}-\phi_n)^3+
(\phi_{n-l}-\phi_n)^3\right]\\
\nonumber
&&-\phi_n^3,
\end{eqnarray}
where $C$ is an arbitrary separation constant.
Its absolute value can be always chosen to be equal
to unity \footnote{The constant $C$ can be removed from equations
(\ref{ODE},\ref{algebra}) by rescaling
$G \rightarrow G/\sqrt{|C|}$ and $\phi_n \rightarrow
\sqrt{|C|}\phi_n$. Note, that this rescaling does not
affect the initial variables $u_n=G\phi_n$.}.

While the dynamics of all the sites is governed by a unique
function $G(t)$, which can be easily found
by integrating Eq.~(\ref{ODE}), the spatial profile of possible
solutions of Eqs.~(\ref{mot}) is determined by Eqs.~(\ref{algebra})
being of main interest for us. 

\section{Basic types of compact-like Discrete Breathers}
\label{sec:three}

It is important to note, that the dynamics of a DB,
as well as its spatial profile, depend on the sign of $C$ in Eqs.~(\ref{ODE},\ref{algebra}).
This sign is fixed only for the case of uncoupled
oscillators ($K=0$) and for small values of $K$ ($C < 0$), while generally it can be arbitrary.
As a consequence, for large enough values of $K$ Eqs.~(\ref{mot}) support
\emph{two classes} of DBs with different dynamical properties, since the sign of $C$
defines the type of nonlinearity ('soft' or 'hard') in the Eq.~(\ref{ODE})
for the master function $G(t)$. DB solutions of different classes possess essentially 
different core structures. As a consequence, the impact of long-range interactions on DB tail structure
is also different for these two classes of DBs.

\subsection{Core structure of DBs}

In order to understand how two different classes of solutions of Eqs.~(\ref{mot}) appear,
it is instructive to start with a simple case of three coupled oscillators. This simple model
gives a rather good approximation for the sites of a DB core,
which are practically not affected by the presence of long-range interactions.
Here we restrict ourselves in considering only symmetric DBs
centered at site $n=0$  (it is straightforward to modify 
this approximate model for more complicated types of 
\emph{multi-site} DBs). Therefore, by putting $\phi_{-1}=\phi_1$
and $\phi_1=\kappa\phi_0$, we finally obtain from Eqs.~(\ref{algebra})
the following expression for the central site amplitude:
\begin{equation}
\label{phi0}
\phi_0^2=-C\frac{1+2\kappa}{1+2\kappa^3\left(1-K\right)},
\end{equation}
while the coefficient $\kappa$ is a root of the fourth-order polynomial
equation
\begin{equation}
\label{polynom}
2K\kappa^4-\left(4K+1\right)\kappa^3+3K\kappa^2+(K+1)\kappa-K=0.
\end{equation}
For each given real value of $\kappa$ obtained from Eq.~(\ref{polynom}) the sign of $C$
is fixed and determined by the condition of non-negative valued r.h.s. of Eq.~(\ref{phi0}).

\begin{figure}
\includegraphics[angle=270, width=0.45\textwidth]{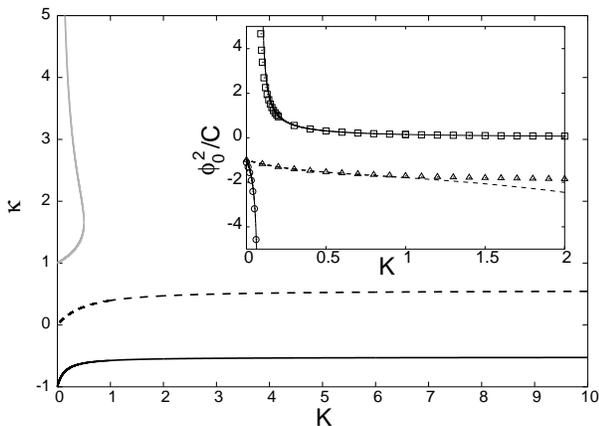}
\caption{Real roots of Eq.~(\ref{polynom}). The inset shows the corresponding values of $\phi_0^2/C$ computed from
Eq.~(\ref{phi0}), squares, triangles and circles indicate the same quantity computed for the
system of 201 coupled oscillators.
}
\label{fig0}
\end{figure}

\begin{figure*}
\includegraphics[angle=270, width=0.32\textwidth]{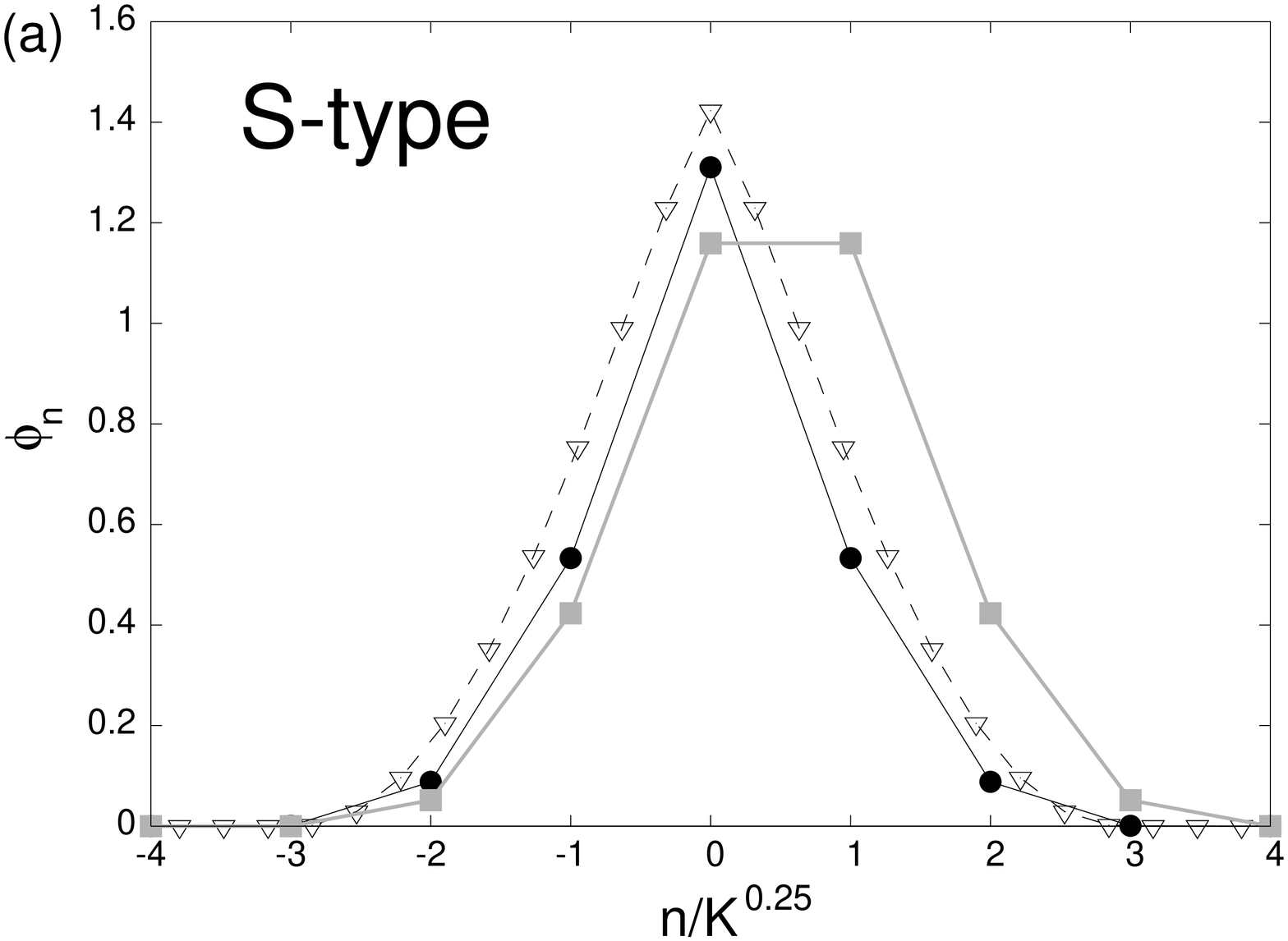}
\includegraphics[angle=270, width=0.32\textwidth]{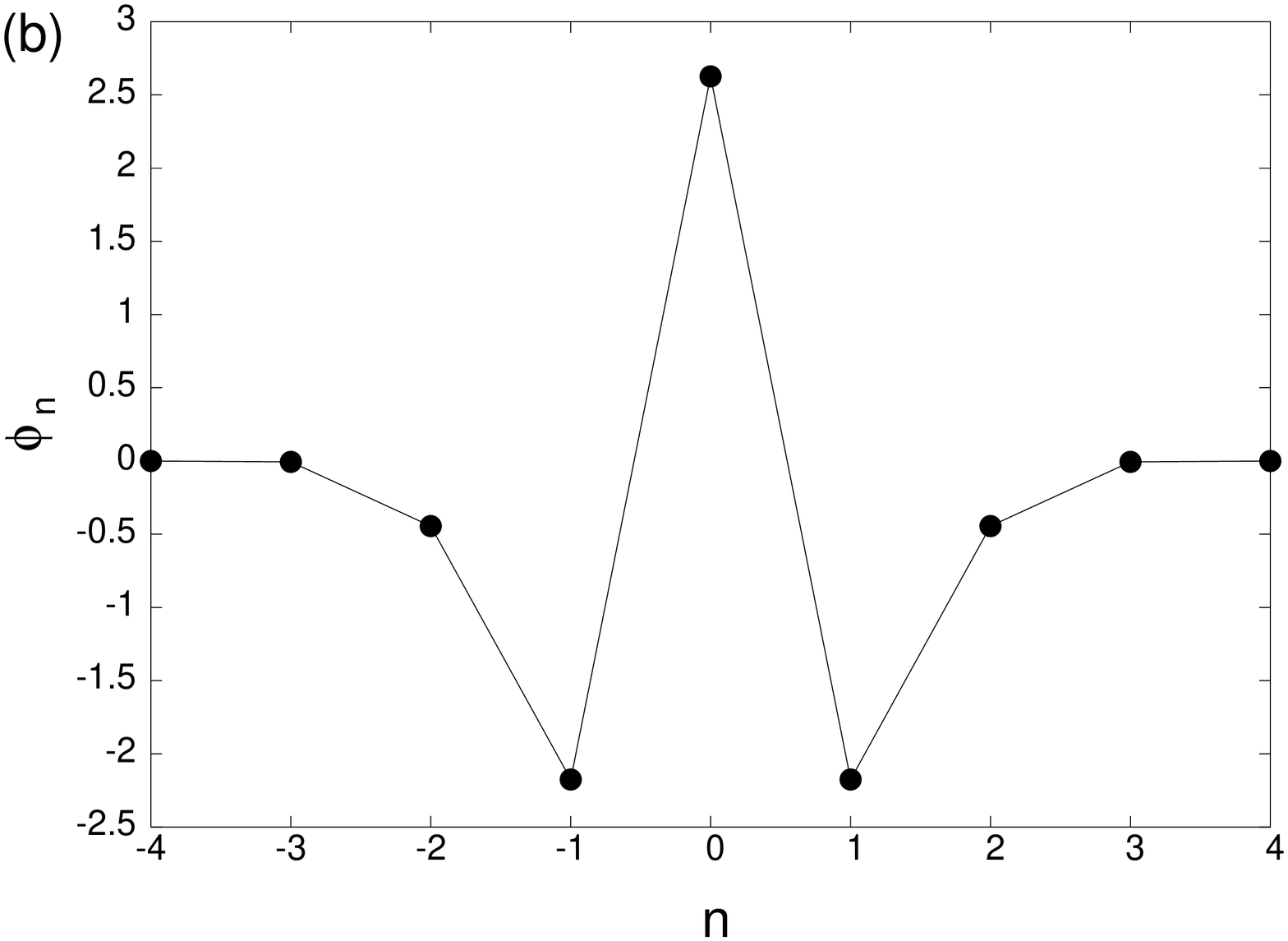}
\includegraphics[angle=270, width=0.32\textwidth]{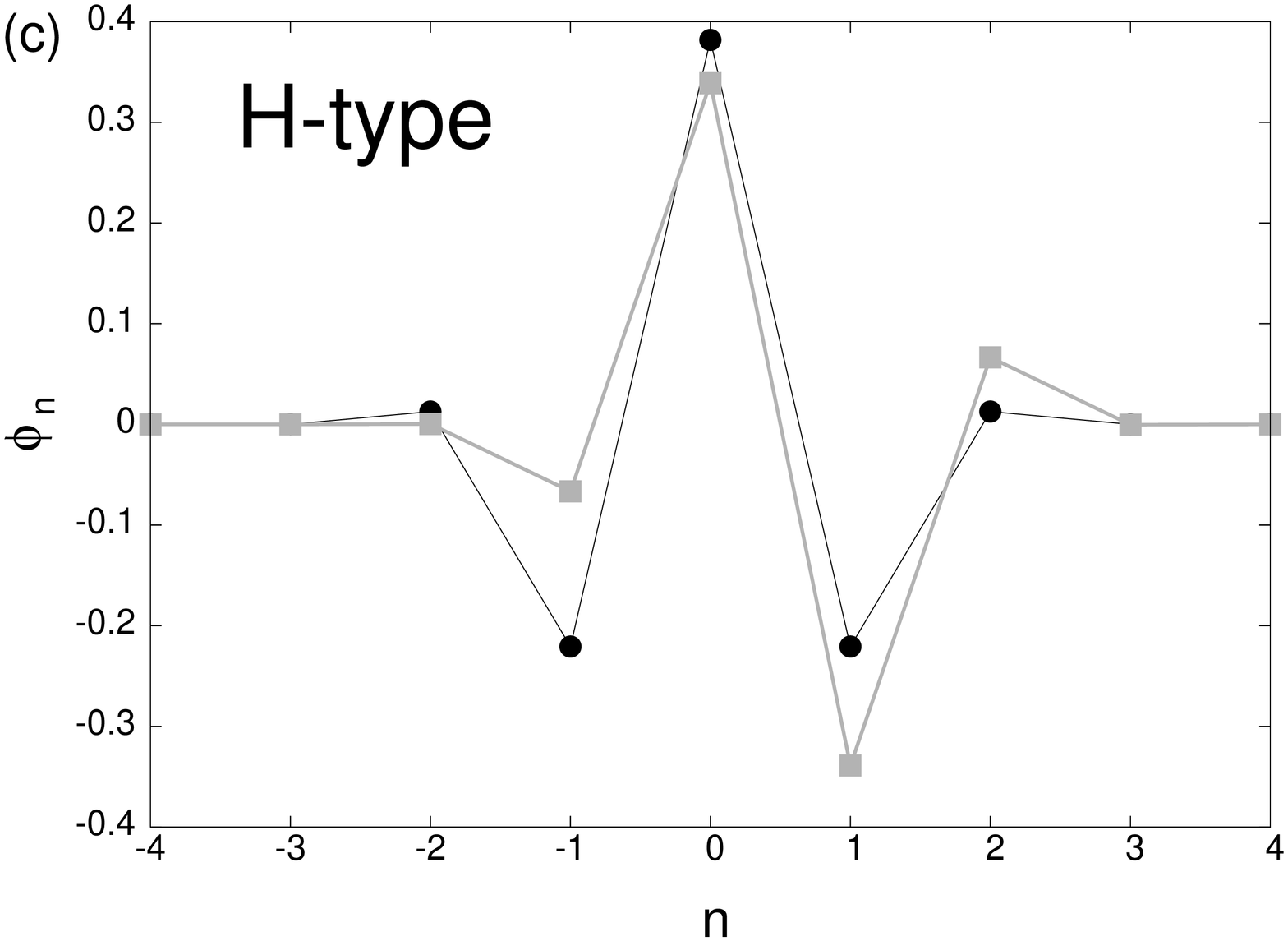}
\caption{Profiles of various types of single-site (black color) and two-site (gray color)
DB solutions of Eq.~(\ref{algebra}):
(a) nonstaggered DBs
corresponding to the branch marked with triangles in the inset in Fig.~\ref{fig0} (S-type DBs), parameter values are: $C=-1$, $K=1$ (circles, squares) and $K=100$ (triangles);
(b) a staggered-core DB corresponding to the branch marked with circles in the inset in Fig.~\ref{fig0}, parameter values are: $C=-1$, $K=0.07$ ($K<K_{cr}$);
(c) staggered-core DBs corresponding to the branch marked with squares in the inset in Fig.~\ref{fig0} (H-type DBs), parameter values are: $C=1$, $K=1$ ($K>K_{cr}$).
The long-range decay rate is $s=100$,
lines are provided to guide the eye.
}
\label{fig:profs}
\end{figure*}

In Fig.~\ref{fig0} the real roots of Eq.~(\ref{polynom}) are plotted
in the range of the coupling constant $K\in [0,10]$. There exist
two real non-trivial roots, whose absolute values stay below
unity [and thus corresponding to single-site DBs with amplitudes
of the central site $\phi_0$ {\it greater} than those of the
neighboring sites $\phi_{\pm1}$] in the whole interval of non-negative
values of $K$ up to $K\rightarrow +\infty$. One of these roots originates
from $\kappa=0$ in the uncoupled limit (i.e. from a single-site excitation)
and remains to be positive in the whole interval of $K$, see dashed
line in Fig.~\ref{fig0}. The corresponding family of DBs have a
non-staggered pattern of amplitudes
$\phi_n$ -- {\it in-phase} oscillations, see Fig.~\ref{fig:profs}(a). The value of $\phi_0^2/C$ computed
for this root from Eq.~(\ref{phi0}) is negative for $K\le 2$ (see dashed line
in the inset in Fig.~\ref{fig0}), therefore one should choose $C=-1$ for
this type of DBs in Eqs.~(\ref{ODE},\ref{algebra}). Thus, DB solutions
of Eq.~(\ref{mot}) with a non-staggered pattern should possess {\it soft}
nonlinear properties, i.e. their amplitude decays with growing frequency,
as it follows from the Eq.~(\ref{ODE}) with negative $C$.
Here we mention, that the quantity $\phi_0^2/C$ computed for the 3-site model
changes its sign above the value $K=2$. However, the comparison with numerically 
obtained solutions
of Eq.~(\ref{algebra}) for a larger system size ($N=201$) indicates essential 
discrepancies when
$K\gtrsim 1$ (see triangles in the inset in Fig.~\ref{fig0}). It comes from the 
fact that a DB core
of breathers with the non-staggered profile extends while increasing $K$, approaching 
continuum compacton solutions \cite{RS05} as $K\rightarrow +\infty$. 
Indeed, as shown in Fig.~\ref{fig:profs}(a), while a DB core involves more and 
more sites with increasing $K$, its characteristic width in terms of the continuum 
coordinate $x=n/K^{0.25}$ remains to be fixed and demonstrates rather good agreement 
with the value $L_0\approx2.92$ reported for DBs in the continuum limit \cite{RS05}.

Another non-trivial real-valued root of Eq.~(\ref{polynom}) originates from $\kappa=-1$ in the 
uncoupled limit (i.e. from a staggered homogeneous excitation) and remains to be negative in
the whole interval of $K$, see solid black line in Fig.~\ref{fig0}. The corresponding DB family
is characterized by a staggered profile in the core (while the tails have more complicated profile, as will be
shown below), i.e. the central ($n=0$) and the neighboring ($n=\pm 1$) sites oscillate in 
{\it anti-phase}. It is remarkable, that the corresponding value of $\phi_0^2/C$ changes its
sign at $K=K_{cr}\approx 0.1$,
see solid black lines in the inset in Fig.~\ref{fig0}. Therefore, for small enough values
of the coupling constant $K<K_{cr}$ both staggered and non-staggered types of DB
solutions of (\ref{mot}) possess {\it soft} nonlinear properties, in accordance to the chosen
type of the on-site nonlinear potential (\ref{onsite}). The profile of the corresponding staggered-core DB solution at $K=0.07$ is shown in Fig.~\ref{fig:profs}(b).
Notably, all the tail sites perform inphase oscillations in this type of DBs, 
similar to the case of nonstaggered DBs described above. However, the central 
site oscillates in anti-phase with all the rest of the lattice.

The competition
between on- and inter-site nonlinearities results in the change of dynamical behavior of the
DBs with staggered core profile as the coupling becomes strong enough: The nonlinear term in
Eq.~(\ref{ODE}) for the master function $G(t)$ becomes of the {\it hard} type for staggered
DBs when $K>K_{cr}$. As the result, the amplitude of such a DB solution increases with growing frequency. The corresponding profile is shown in Fig.~\ref{fig:profs}(c). In this type of staggered DBs all sites in the core perform anti-phase oscillations. In the case of pure short-range interactions in the system, the staggered pattern persists for the whole spatial profile, including the breather tails. However, long-range interactions destroy the uniform staggered pattern, introducing a complicated domain-like structure in the breather tails, as will be explained below.

Unlike DBs with non-staggered profile, the staggered core DBs stay localized on a
few sites as $K$ increases, therefore the 3-site model gives a rather good approximation for
larger size systems for arbitrarily large values of $K$ (see squares in the inset in Fig.~\ref{fig0}).
Simultaneously, the amplitude of these DBs decreases as $K\rightarrow +\infty$ (for a fixed
value of the frequency). However, we note that there is no limit in frequency (and therefore, in energy) 
for this type of DBs, since the
master function $G(t)$ satisfies the Duffing equation (\ref{ODE})
with the \emph{hard} nonlinear term ($C=1$). Note also, that as $K\rightarrow +\infty$, the influence of the
inter-site nonlinear interactions becomes more important than the effect of on-site nonlinearities for the
oscillating in anti-phase sites of the DB core. Therefore, at large values of $K$ the staggered-core DB asymptotically 
approaches the high-energy limit of discrete breather solutions in models with purely inter-site nonlinearities 
(Fermi-Pasta-Ulam lattices) \cite{weFPU}. 

Thus, for $K>K_{cr}$ Eqs.~(\ref{mot}) support \emph{two different classes} 
of DB solutions with \emph{soft} and
\emph{hard} nonlinear properties, in what follows we will refer to such breather solutions as \emph{S-type} and
\emph{H-type} DBs, respectively. The above discussed single-site DBs represent only particular (basic) 
members of these two classes of solutions. In general, one can constract more complicated localized S- and H-type
solutions - \emph{multi-site} DBs. As an example, we mention here \emph{two-site} DBs with two sites in the core oscillating with the same (maximum) amplitude, see gray lines and symbols in Fig.~\ref{fig:profs}. The center of energy density distribution is located in-between two sites in these DBs, so that they can be viewed as translated half site single-site DBs.


Finally, we would like to remark, that the existence of the two different classes
of DB solutions of Eqs.~(\ref{mot}) is the result of a competition between soft
nonlinearity of the on-site potential and hard nonlinear inter-site interactions.
Upon change the type of nonlinearity in the on-site potential [by changing the
sign of the quartic term in Eq.~(\ref{onsite})], only the H-type DBs with
staggered core profile survive.

\subsection{Linear stability}

Different dynamical properties of DBs with staggered and non-staggered core
profiles result, in particular, in different stability properties of these
excitations. 
In this section we perform linear stability analysis of basic types of $H-$ and $S-$ type DBs
by
studying the dynamical behavior of a small perturbation $\epsilon_n(t)$
to a given DB solution $\hat{u}_n(t)$. In order to construct a certain type of
DB solution of Eq.~(\ref{mot}), we solve numerically Eqs.~(\ref{algebra}) for the
DB profile $\hat{\phi}_n(t)$ with a certain sign of the separation constant $C$
(chosen in accordance to the above performed analysis of DBs structure) and
multiply it by a periodic solution $\hat{G}(t)$ of the Duffing equation
(\ref{ODE}), according to the ansatz (\ref{time-space}). Because of the
time-space separation (\ref{time-space}), the stability properties remain
qualitatively the same for the whole family of a given DB type with different
frequencies $\Omega_B=2\pi/T_B$ [i.e. with different time-periodic
functions $G(t+T_B)=G(t)$]. However, they might drastically change by varying
the relative strength of the on-site and inter-site nonlinearities controlled by
the coupling constant $K$.

Thus, once a given DB solution $\hat{u}_n(t)$ is obtained numerically, we
add a small perturbation
to it $u_n(t)=\hat{u}_n(t)+\epsilon_n(t)$ and linearize equations of
motion Eqs.~(\ref{mot}) with respect to
$\epsilon_n(t)$:
\begin{eqnarray}
\label{linearized}
\ddot{\epsilon}_n&=&3K\sum_{l}\frac{1}{l^s}\left\{\left(\hat{u}_{n+l}-\hat{u}_n\right)^2
\left(\epsilon_{n+l}-\epsilon_n\right)+\right.\\
\nonumber
& &\left.\left(\hat{u}_{n-l}-\hat{u}_n\right)^2
\left(\epsilon_{n-l}-\epsilon_n\right)\right\}-\epsilon_n + 3\hat{u}_n^2\epsilon_n\;.
\end{eqnarray}
In order to simplify the stability analysis, here we restrict ourselves in 
considering pure short-range
interaction terms, thus we keep only the term with $l=1$ in the r.h.s. of
Eqs.~(\ref{linearized}), which corresponds to the limit $s \rightarrow \infty$. 
Being essential for DB tail characteristics, long-range 
interactions practically do not
affect the core structure of a DB, provided that their decay rate $s$ is sufficiently large 
($s>3$). Therefore, the impact of long-range 
interactions on stability properties of DBs 
is expected to be negligible.

The discrete breather acts as a parametic driver for small perturbations
$\epsilon_n(t)$ with the period $T=T_B/2$ being the half-period of the DB
solution [i.e. the half period of a given solution $G(t)$ in Eq.~(\ref{ODE})].

Equations (\ref{linearized}) define a map
\begin{equation}
\left(
\begin{array}{c}
\vec{\dot{\epsilon}} (T) \\ \vec{\epsilon} (T)
\end{array}
\right) = {\cal F}
\left(
\begin{array}{c}
\vec{\dot{\epsilon}} (0) \\ \vec{\epsilon} (0)
\end{array}
\right)
\label{map}
\end{equation}
which maps the phase space of perturbations onto itself
by integrating each point over the period $T$.
Here we used the abbreviation $\vec{x}\equiv (x_1,x_2,...,x_l,...)$.
The map (\ref{map}) is characterized by a symplectic Floquet matrix ${\cal F}$,
whose complex eigenvalues $\lambda$ and eigenvectors $\vec{y}$ provide
information about the stability of the DB \cite{DB_rev}.
Here we note that if all eigenvalues $\lambda$
are by modulus one, then the DB is linearly (marginally) {\it stable}.
Otherwise perturbations exist which will grow in time (typically
exponentially) and correspond to a linearly {\it unstable} DB.
Upon changing a control parameter (e.g. the coupling constant $K$) stable DBs
can become unstable (and vice versa). Such a change of stability is
appearing because two (or more) Floquet eigenvalues collide on the unit
circle and depart from it \cite{DB_rev}.

\begin{figure}
\includegraphics[angle=270, width=0.45\textwidth]{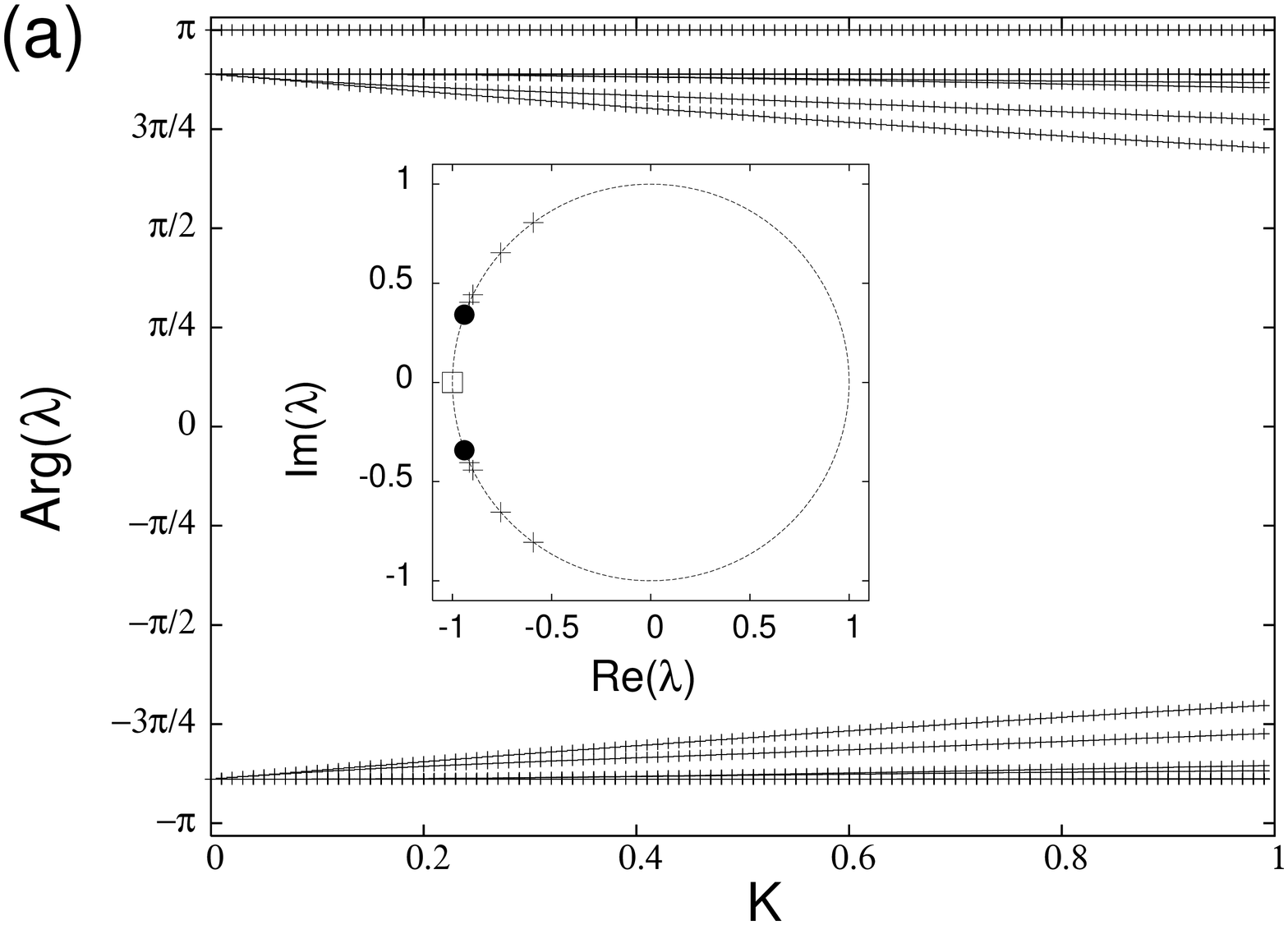}
\includegraphics[angle=270, width=0.45\textwidth]{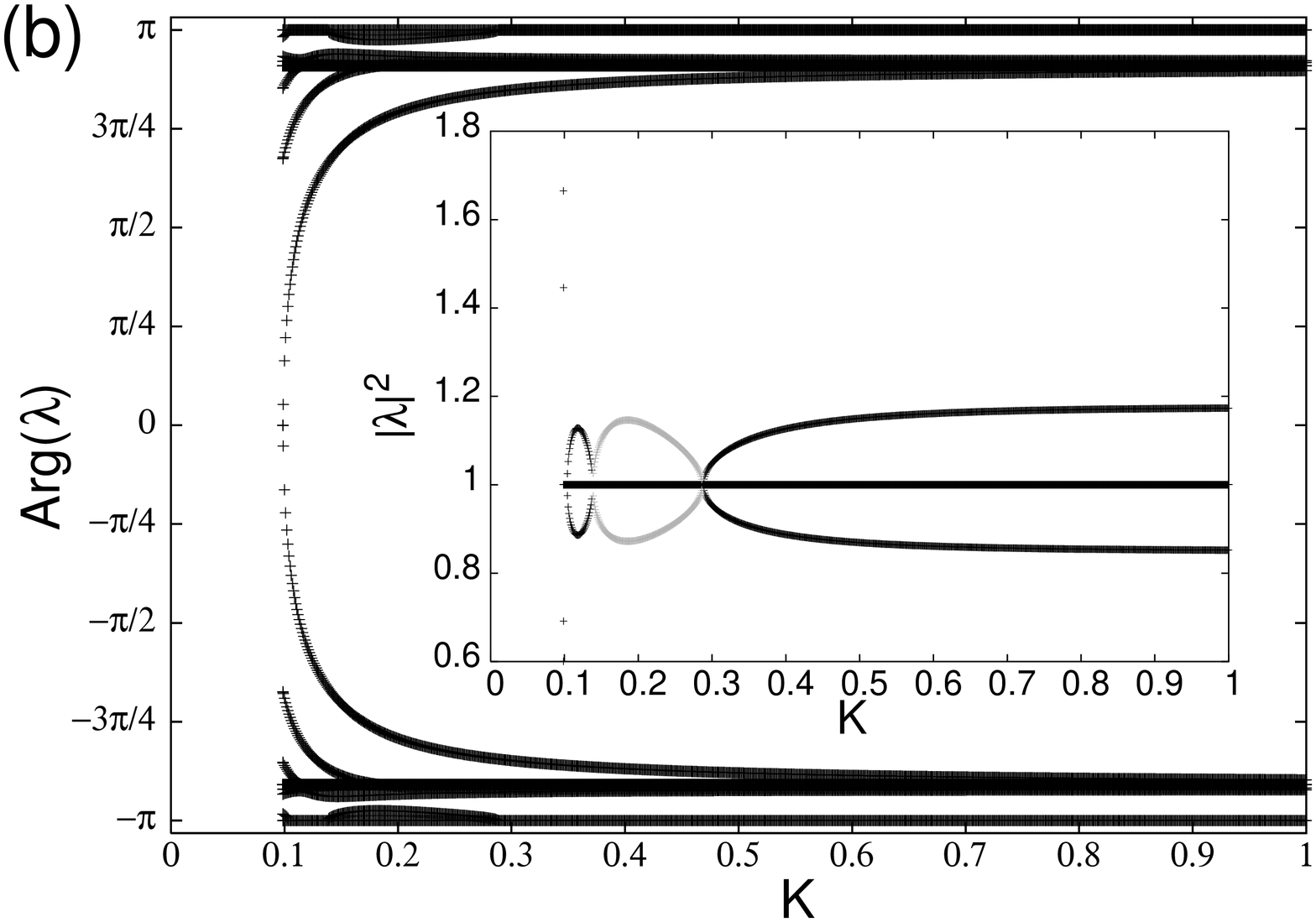}
\caption{Evolution of the arguments of Floquet eigenvalues $\lambda$ with
change of the coupling constant $K$
for different types of single-site DBs: (a) S-type DB with the frequency $\Omega_B=0.9$. Inset shows the
Floquet spectrum at $K=1$, the unit circle in the complex plane is indicated
to guide the eye; (b) H-type DB with the frequency $\Omega_B=1.1$. Inset shows evolution of absolute values $|\lambda|^2$
with change of $K$. A pair of eigenvalues corresponding to the unstable depinning mode (see the main body text for the details) of the two-site DB is indicated with gray color.
}
\label{fig:stab}
\end{figure}

In Fig.~\ref{fig:stab}(a) the typical Floquet spectrum is shown for a
\emph{S-type} single-site DB.
All the eigenvalues
$\lambda$ can be divided into three sub-categories: One pair of eigenvalues is always situated at 
$\exp(i\pi)$ [denoted by
squares in the inset in Fig.~\ref{fig:stab}(a)]; it corresponds to perturbations
along the DB periodic orbit (phase mode) and along the corresponding family of
DB solutions \cite{DB_rev}. The period of these perturbations coincides with the
DB period.
In addition, there are quasi-degenerated~\footnote{The eigenvalues are slightly
different due to the noncompact breather profile.} 
bands of eigenvalues at $\lambda=\exp(\pm iT_B/2)$ [filled
circles in the inset in Fig.~\ref{fig:stab}(a)], which correspond to
perturbations in breather tails. Such perturbations have characteristic
frequency
$\omega_{l}=1$ defined by the choice of the linear constant in the
on-site potential (\ref{onsite}), they would correspond to linear phonons if 
linear coupling between sites were introduced. Finally, there is a finite number
of eigenvalue pairs bifurcating from the quasi-degenerated bands, which correspond
to perturbations of the DB core sites [crosses in the inset in
Fig.~\ref{fig:stab}(a)]. The number of such isolated pairs is proportional to the characteristic DB core
size, it grows as the coupling constant $K$ increases, see
Fig.~\ref{fig:stab}(a). While increasing the coupling constant $K$, these
isolated pairs move on the unit circle and new pairs bifurcate from the
quasi-degenerated band, but they do not collide with each other and the
non-staggered type of DB remains linearly stable up to the continuum limit
$K\rightarrow +\infty$.

Notably, the Floquet spectrum of the S-type two-site DB is qualitatively the same as the one of the single-site DB for any
$K$. The only principal difference is that it has two degenerated pairs of eigenvalues at $\lambda=\exp(i\pi)$, since in the 
uncoupled limit $K=0$ the corresponding solution has two sites excited with equal amplitude. Usually such degeneracy of eigenvalues is lifted for any nonzero $K$, and one pair of eigenvalues is "pushed out" from $\lambda=\exp(i\pi)$ either along the real axis or along the unit circle. However, in the case of purely nonlinear interactions between sites the \emph{symmetric} two-site DB "cuts"
the effective linear chain (\ref{linearized}) into two non-interacting halves. Therefore, the additional degeneracy of 
eigenvalues corresponding to symmetric and anti-symmetric perturbations with respect to the DB center remains for any $K$.

In contrast to the S-type DBs, the H-type single-site and two-site DBs 
are linearly stable only within
certain windows of the coupling constant $K$ values, see Fig.~\ref{fig:stab}(b).
Close to the critical value of the coupling constant $K=K_{cr}$, below which the H-type DBs do not exist, 
both single- and two-site H-type DBs experience strong instabilities connected with tangent bifurcations of these
solutions with other ones, having more complicated spatial structure. In addition, there is another instability of a finite
strength, appearing in certain windows of the parameter $K$, see inset in Fig.~\ref{fig:stab}(b).
In general, apart from several small intervals in $K$, for any given value of $K$ only one of these two DB configurations is stable. The corresponding unstable perturbation -- the "depinning" mode -- "tilts" the single-site (two-site) DB towards the half-site shifted stable two-site (single-site) one.
Changing the coupling constant, the stable configuration varies from the two-site to single-site DB and back [at $K\approx 0.15$ and $K\approx 0.3$, see Fig.~\ref{fig:stab}(b)], so that
the \emph{exchange of stability} process \cite{stab_inv} is observed. 
This exchange of stability process can be connected to an exchange of the dominant roles between inter-site and on-site nonlinearities.
Indeed, typically for models with purely inter-site nonlinearities (Fermi-Pasta-Ulam lattices) the basic stable configuration is the two-site DB \cite{DB_rev, weFPU}. In contrast, for models with weak coupling between sites and a non-linear onsite potential in the form (\ref{onsite})
(Klein-Gordon lattices) the single-site DB configuration is stable \cite{DB_rev}.

\begin{figure*}
\includegraphics[angle=270, width=0.32\textwidth]{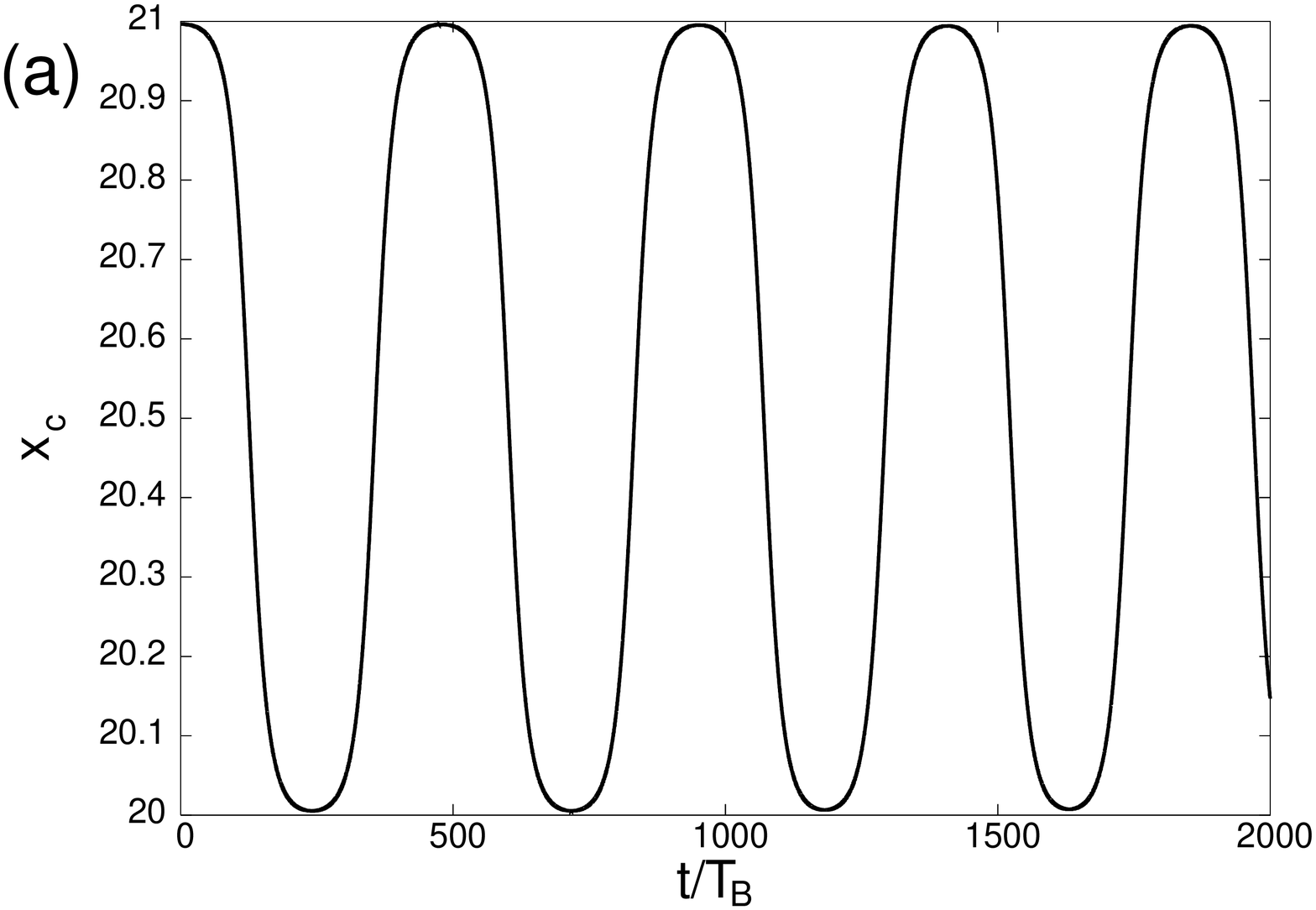}
\includegraphics[angle=270, width=0.32\textwidth]{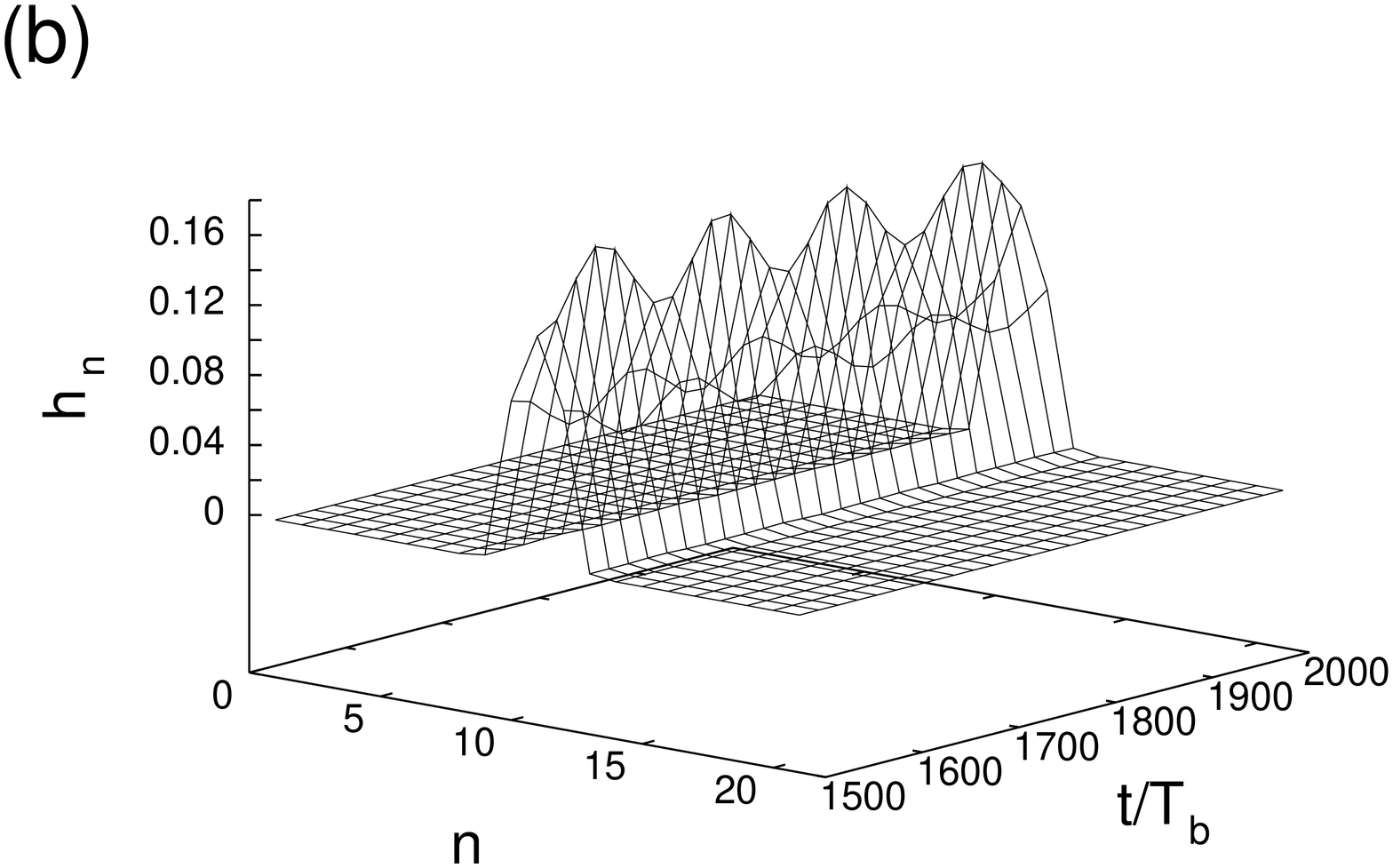}
\includegraphics[angle=270, width=0.32\textwidth]{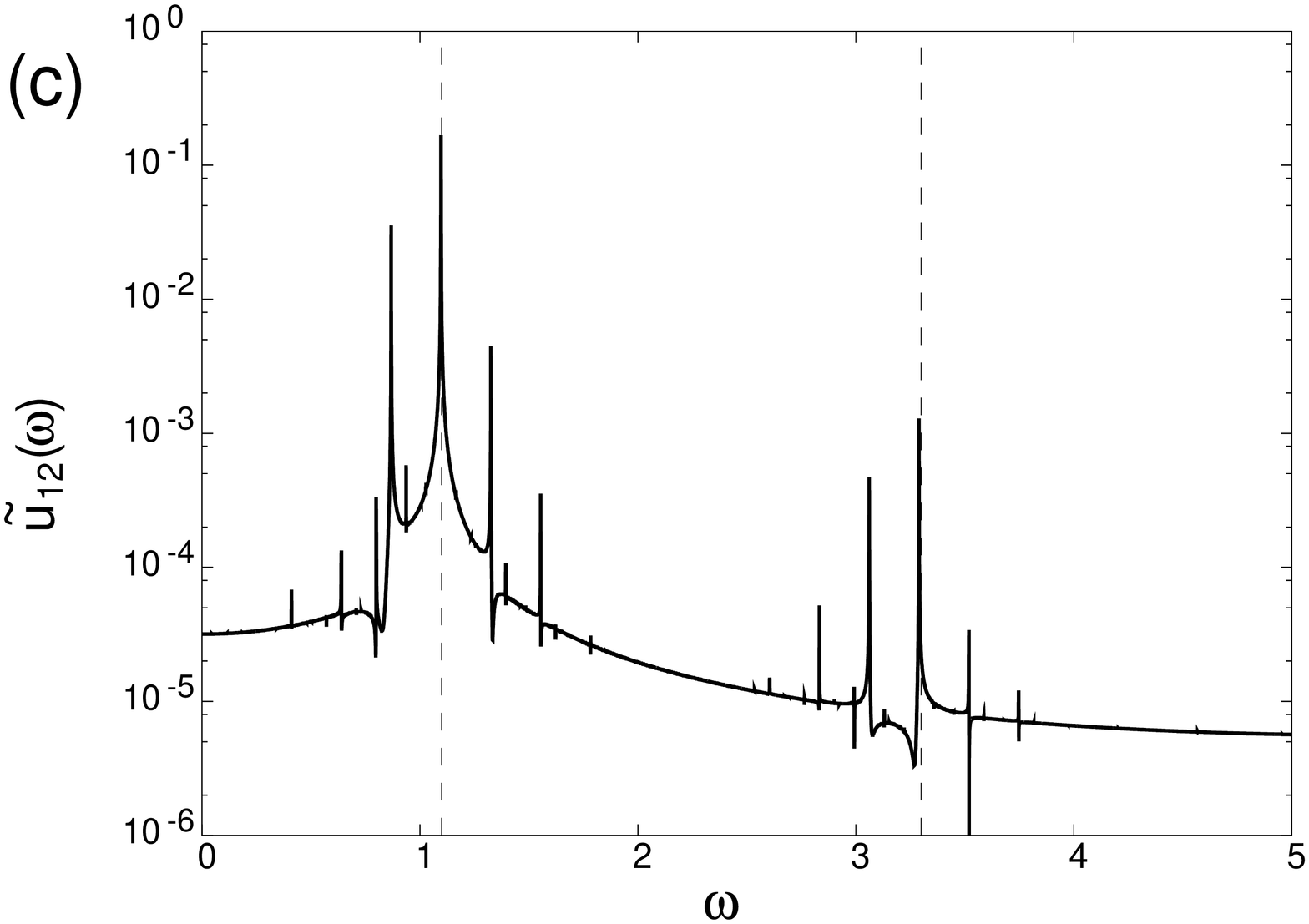}
\caption{
Dynamics of a perturbed DB:
(a) Position of the DB center $x_c$, calculated through the energy
dencity $h_n$ (\ref{hamilt}): $x_c=\sum_n (n \cdot h_n)/\sum_n h_n$. 
At time $t=0$ the single-site H-type DB ($\Omega_B=1.1$, $K=0.3$) 
was perturbed along the unstable depinning mode with the relative energy of 
perturbation $\Delta H/H_{DB}\sim 10^{-4}$;
(b) Energy density $h_n$ (\ref{hamilt}) evolution in dynamics of the stable single-site 
H-type DB ($\Omega_B=1.1$, $K=0.2$) with a symmetric perturbation along one of the DB internal modes. 
The relative energy of 
perturbation $\Delta H/H_{DB}\sim 10^{-2}$; (c) The Fourier transform of $u_{12}(t)$ (the DB is centered at $n=11$) 
for the same dynamical simulation as in (b). Vertical dashed lines indicates the first and the third DB harmonics $\Omega_B=1.1$ and
$3\Omega_B=3.3$.
}
\label{fig:oscil}
\end{figure*}

Of principal interest is an influence of the unstable depinning mode on the
dynamical behavior of staggered-core DBs. A small perturbation along this mode
is generally known to result in depinning of the unstable DB
from its initial position. Depending on the relative Hamiltonian energy (\ref{hamilt}) of the perturbation
($\Delta H/H_{DB}$) and on the strength of instability of the depinning mode,
the resulting behavior might vary from quasiperiodic-like oscillations between
two neighboring unstable positions (in a well of the corresponding
"Peierls-Nabarro potential") to quasi-regular or even chaotic-like motion along
the chain (see e.g. \cite{stab_inv, Chen}). Generally, the depinned DB resonates with
linear phonons through its excited internal modes and starts to radiate energy.
Therefore, eventually it will be trapped again at some stable position or even
disappear completely transferring totally its energy to excited delocalized
phonons.

However, in the case of purely nonlinear
dispersion there are no linear phonons in the system, and all possible
linear resonances are suppressed. As a result, one can observe almost perfect
quasi-periodic oscillations of perturbed unstable DBs between two neighboring stable positions, 
see Fig.~\ref{fig:oscil}(a). 
Similar quasi-periodic behavior is observed in dynamics of a stable DB with perturbation 
along one of its internal modes, see Fig.~\ref{fig:oscil}(b,c). 
In the latter case no detectable radiation (within the used double precision) 
of energy from the DB core was observed
during dynamical simulation over $10^5$ breather periods.
In this respect, a
one-dimensional Hamiltonian lattice with purely anharmonic 
interactions between sites (\ref{hamilt}) might be an interesting "toy model" to study
more complicated \emph{exact} DB solutions like quasi-periodic and moving
discrete breathers.

\section{Long-range interactions effect}
\label{sec:four}

\begin{figure}
\includegraphics[angle=270, width=0.45\textwidth]{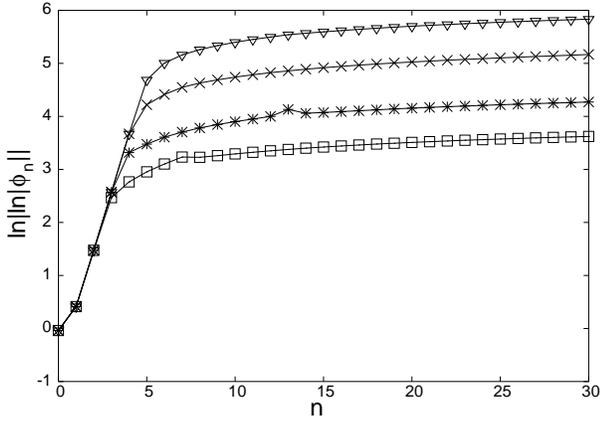}
\caption{Single-site H-type DB solutions of
(\ref{algebra}) with $C=1$ and $K=1$ for various exponents of the long-range interaction:
$s=10$ (squares), $s=20$ (stars), $s=50$ (crosses) and $s=100$
(triangles).
}
\label{fig1g}
\end{figure}

Let us now fix the value of $K=1$ and study the influence of long-range
interactions on the spatial profile of a DB.
In Fig.~\ref{fig1g} the profiles of the H-type single-site DBs are
shown for various values of the decay constant $s$. They
were obtained by solving Eqs.~(\ref{algebra}) numerically
with use of the standard
Newton scheme \cite{Fla04} for the chain of
$N=201$ oscillators ($-100<n<100$) with periodic boundary
conditions~\footnote{While using the periodic boundary
conditions, we introduce a cutoff for the interaction
range exactly at one half of the system size, to avoid
double counting of interactions.}. Similarly to
the case of nonlinear short-range interaction, the computed DBs
have a compact-like structure with mainly three central sites
oscillating (see Fig.~\ref{fig:profs}).
The
other oscillators are almost at rest and can be considered as
breather tails.
The presence of long-range interactions breaks
the uniform super-exponential law of the spatial tail decay known for the case of pure short-range nonlinear interactions \cite{Fla94, DEF+02},
introducing several cross-over lengths.

For a few central sites, lying within the breather core,
the forces due to the
long-range interactions are negligible as compared to those due
to the nearest-neighbor interactions. Thus, the central part of
a breather is practically not affected by the presence of long-range
interactions. However, at some distance $L_1$ from the DB center
interactions with the nearest neighbors (having small enough amplitudes)
become of the same order as the long-range interactions with the DB core
(central three sites having the highest amplitudes). This distance
is the first cross-over length, where the long-range interactions
come into play. It can be roughly estimated by an assumption,
that at the distance $L_1$ interactions with the breather core
sites are exactly compensated by interactions with the nearest
neighbors. Thus, keeping only the leading order terms in the
sum in the r.h.s. of (\ref{algebra}) one obtains:
\begin{equation}
\label{L1_estimate}
-\frac{\phi_1^3}{(L_1-1)^s}=\phi_{L_1-1}^3.
\end{equation}
Since for $|n|<L_1$ the relations between the amplitudes $\phi_n$ are
practically the same as in the case of pure nearest-neighbor interactions,
i.e. they follow the super-exponential law
$\phi_n \sim (-1)^n |\phi_1|^{\exp(\ln(3)|n-1|)}, 1\le |n|\le L_1$,
one can obtain from (\ref{L1_estimate}) :
\begin{equation}
\label{L1_est2}
|\phi_1|^{\exp\left[3\ln3 (L_1-2)\right]-3}(L_1-1)^s = 1.
\end{equation}
In the limit of extremely large values of $s$ the distance $L_1$ will
be also large, and satisfy
\begin{equation}
\label{L1_est_fin}
L_1\sim \frac{\ln s}{\ln 3}+2.
\end{equation}
Thus the
first cross-over length $L_1$ grows approximately logarithmically with $s$.
The numerical results in Fig.~\ref{fig1g} yield
$L_1(s=10)\approx 3$, $L_1(s=20)\approx 4$, $L_1(s=50)\approx 5$
and $L_1(s=100) \approx 6$. They compare very well with the
corresponding solutions of (\ref{L1_est2}): 2.71, 3.46, 4.44, 5.17.
Therefore, even extremely fast (but still algebraically) decaying in space
long-range interactions essentially destroy the concept of
compact-like breathers, since only amplitudes of a few sites
in the tails obey the super-exponential law of decay, while the rest of the tail amplitudes
decay much slower in space
(following a power law, as will be shown below).

\begin{figure}
\includegraphics[angle=270, width=0.4\textwidth]{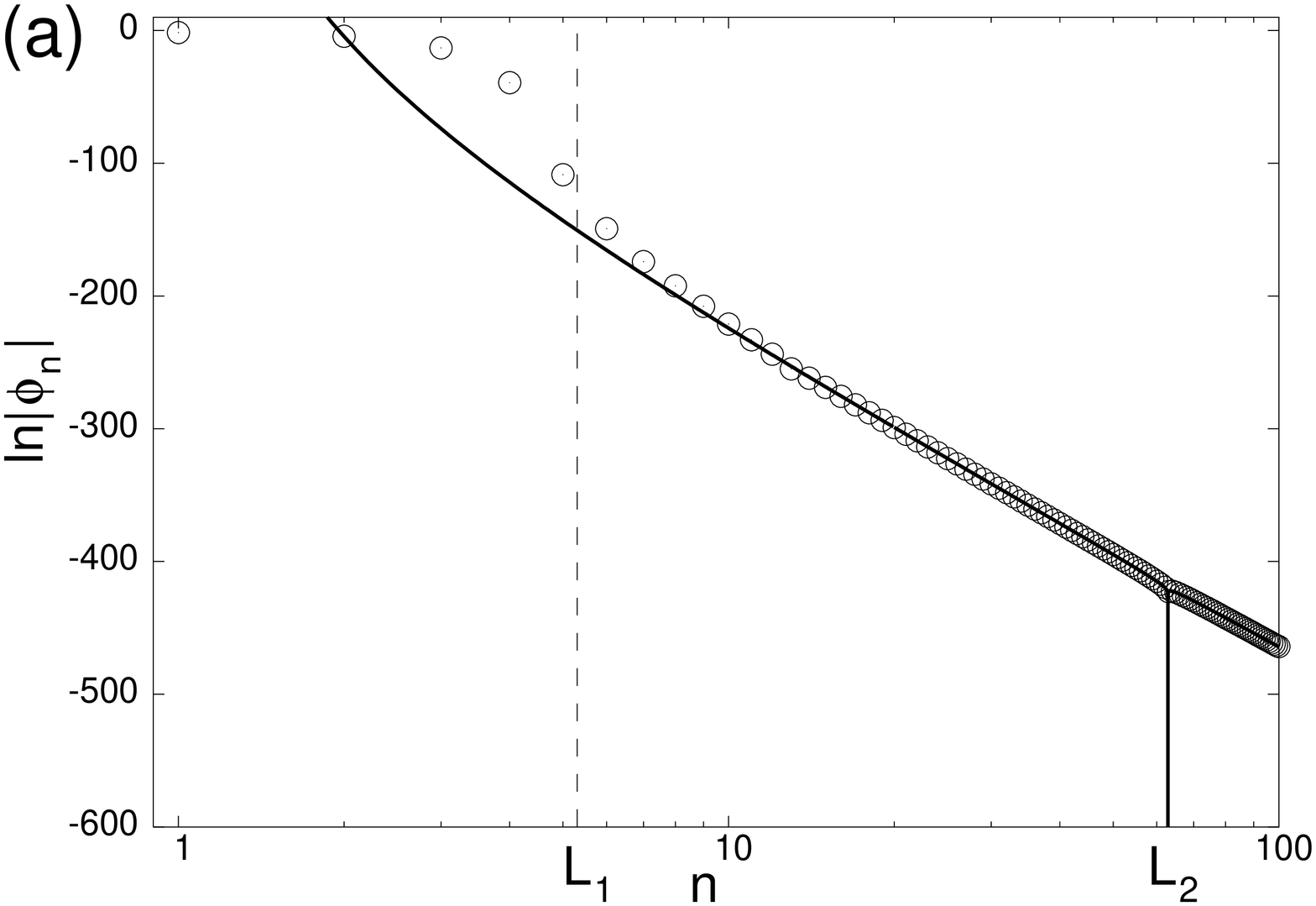}
\includegraphics[angle=270, width=0.4\textwidth]{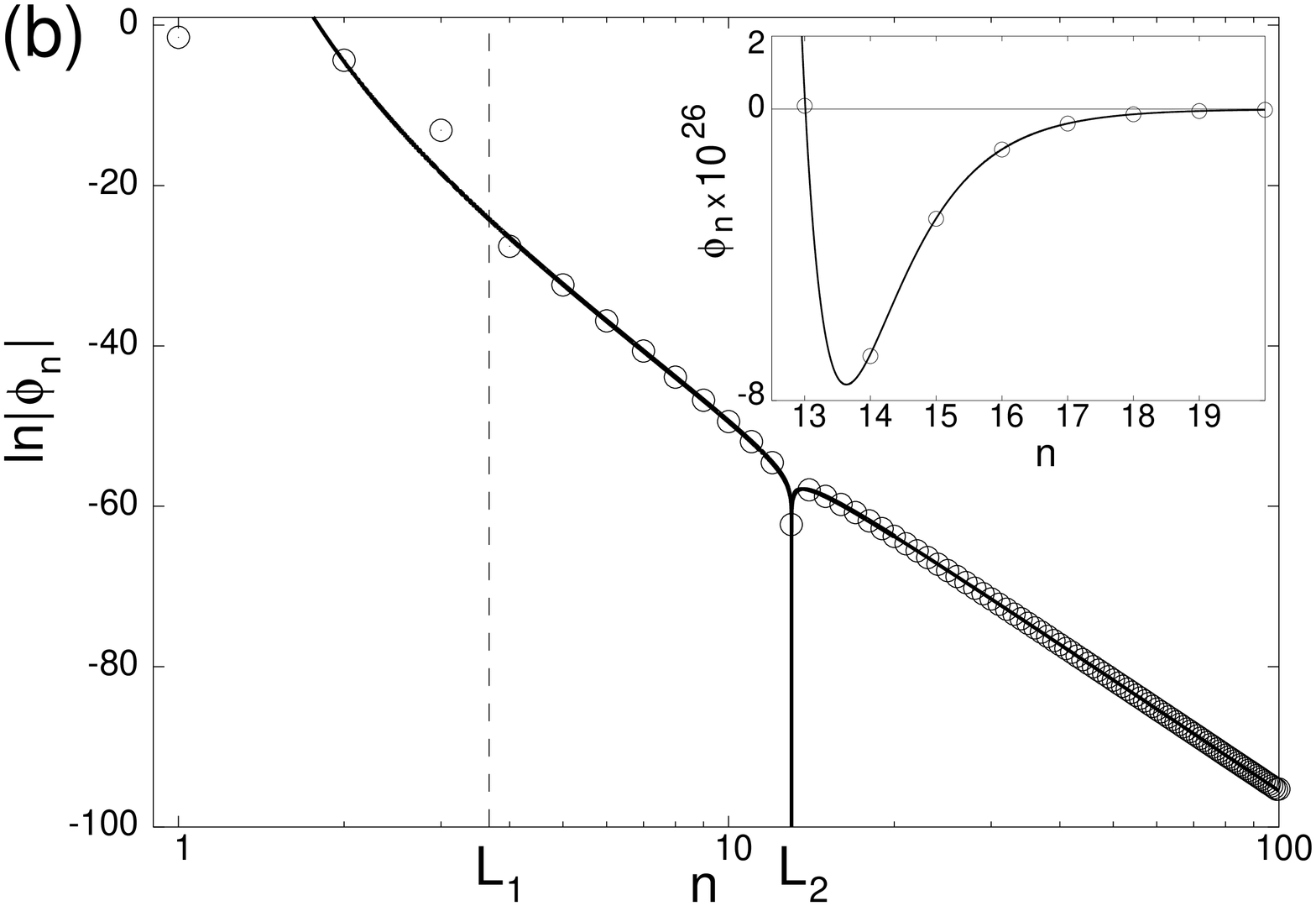}
\caption{Single-site H-type DB solutions of (\ref{algebra}) with $C=1$
for various exponents of the long-range interaction:
(a) $s=100$; (b) $s=20$. Circles: numerical results.
Solid lines: tail asymptotes
(\ref{powerlaw}).
Dashed lines: location of $L_1$.
The inset in (b) indicates the change of amplitudes sign around the
cross-over point $L_2$.}
\label{fig2g}
\end{figure}

At large distances from the breather center $n\gg 1$ (due to
the single-site DB symmetry around $n=0$ we  consider here only
non-negative values of $n$) the impact of short- and
long-range interactions is exchanged: now the most powerful
contribution comes from the interaction with the breather core,
while nearest neighbors, due to their small amplitudes,
practically do not affect the dynamics of a tail site.
Thus, for large $n$ one can derive from (\ref{algebra})
the following asymptote:
\begin{equation}
\label{powerlaw}
\phi_{n}\approx -\frac{K}{C}\left[\frac{\phi_0^3}{n^s}+
\frac{\phi_1^3}{(n-1)^s}+\frac{\phi_1^3}{(n+1)^s}\right],
\end{equation}
which in fact gives a rather good approximation for all
tail sites starting from the first cross-over point $n=L_1$
(see solid lines in Fig.~\ref{fig2g}). Note that only amplitudes of the two DB core sites
and the sign of separation constant $C$ are needed to obtain this asymptote for tail amplitude
distribution. 
In this respect we found the simple three-site model, discussed in Sec.~\ref{sec:three}, to be 
very fruitful: it gives full information not only about the DB core sites, but about tail characteristics as well.

Note that the specific structure of a staggered DB core with a central site $n=0$ and two neighboring sites $n=\pm 1$ having amplitudes $\phi_n$ of opposite signs
stipulates several other cross-over lengths
connected to changes of the sign of the r.h.s. in (\ref{powerlaw})
which manifest as singularities in the
logarithmic plots in Fig.\ref{fig2g}.
The most pronounced cross-over at $n=L_2$ is associated with the
change from a single power law $n^{-s}$ to a more complex
one (\ref{powerlaw}), see Fig.~\ref{fig2g}. Indeed, in the
case $n\gg 1, (s/n)\ll 1$ the expression (\ref{powerlaw}) can
be re-written as:
\begin{equation}
\label{powerlaw_as}
\phi_n\approx -\frac{K \phi_0^3}{C n^s}\left\{
1+2\kappa^3+\kappa^3\frac{s^2}{n^2}+o\left(\frac{s^2}{n^2}\right)
\right\},
\end{equation}
($C=1, \kappa<0$ for a staggered core DB and $C=-1, \kappa>0$ for a non-staggered DB).
Thus, in leading order,
at large enough distance from the DB center $n>L_2$
the tail amplitudes follow the same power law
$n^{-s}$ as the decay of long-range interactions.
Since $L_2$ is defined by the vanishing of the bracket
on the right hand side of Eq.~(\ref{powerlaw_as}) we obtain
in leading order
\begin{equation}
L_2 \approx s \sqrt{\frac{-\kappa^3}{1+2\kappa^3}}\;.
\label{L_2}
\end{equation}
The corresponding values of $L_2$ for $s=10,20,50,100$
and $\kappa=-0.382$
with
(\ref{L_2}) are
$5.6,11.2,28,56$. They compare reasonably well with the numerically
observed ones $L_2(s=10)\approx 7$,
$L_2(s=20)\approx 13$, $L_2(s=50)\approx 31$
and $L_2(s=100) \approx 64$.
In-between the two characteristic length scales $L_1<n<L_2$ the tail
amplitudes decay following a more complicated power law (\ref{powerlaw}).

\section{Conclusions}
\label{sec:concl}

To conclude, we revealed the influence of long-range nonlinear interactions on the spatial 
profile and properties of compact-like discrete breathers in a model of coupled oscillators 
with pure nonlinear dispersion. As we demonstrate, it is the intriguing property of the model 
under consideration, that it supports two classes of discrete breathers -- with 
{\it staggered} and {\it non-staggered} spatial profiles of a DB core -- having different 
dynamical properties. The dynamics of a non-staggered DB is essentially governed by the 
{\it soft} nonlinear on-site potential, while DBs with the staggered core have the opposite, 
{\it hard}, type of nonlinear dynamical behavior caused mainly by the presence of nonlinear 
interactions in the chain. Apart from different dynamical properties,
the influence of long-range interactions on spatial profiles of these two types of DBs is also different.
With the algebraic spatial decay long-range interactions introduce a new length scale
which
becomes essential at large enough distances from a DB core. We show, that
the effect 
of long- and short-range terms competition results in the appearance of a characteristic 
cross-over length $L_1$ in both types of DBs, at which the spatial tail decay drastically 
changes from the super-exponential law to the algebraic one. For large powers $s$ of the long-range 
interactions spatial decay the cross-over length $L_1$ scales logarithmically with $s$. The 
tail asymptote (\ref{powerlaw}) demonstrates complex power law spatial decay, which follows 
essentially the same algebraic decay as the long-range interactions in the system at large enough 
distances from the DB core. While for  non-staggered DBs the influence of long-range interactions 
manifests through the only characteristic cross-over length $L_1$, the spatial profile of DBs 
with the staggered core possess several other cross-over lengths associated with sign changes of 
the asymptote (\ref{powerlaw}). Thus, the spatial pattern of oscillations in a DB with 
the staggered core becomes rather complicated in the presence of long-range interactions: its 
core sites perform anti-phase oscillations, while
its tails are splitted in several domains of in-phase oscillations.

Finally, we would like to mention that the discussed 
case of purely nonlinear coupled oscillators represents a simple model to 
reveal properties
of nonlinear excitations in a system without linear phonons. As we 
demonstrated in this paper, "switching off" the phonons
leads not only to the change in characteristic rate of spatial localization 
of energy, but to appearance of several other
intriguing \emph{dynamical} properties of nonlinear localized excitations. 
Especially it eliminates the possible source of linear resonances which 
otherwise would destroy quasiperiodic breathers and possibly also 
moving breathers. The discussed models thus
allow to obtain
a better understanding
of the general problem of existence/non-existence of quasi-periodic and moving 
discrete breather solutions.

\acknowledgments{
We would like to thank M. Johansson for careful reading of the manuscript and providing us with usefull
comments. We also thank P.~Rosenau, T.~Bountis and P.~Maniadis for usefull and stimulating discussions.
}


\begin{thebibliography}{21}
\expandafter\ifx\csname natexlab\endcsname\relax\def\natexlab#1{#1}\fi
\expandafter\ifx\csname bibnamefont\endcsname\relax
  \def\bibnamefont#1{#1}\fi
\expandafter\ifx\csname bibfnamefont\endcsname\relax
  \def\bibfnamefont#1{#1}\fi
\expandafter\ifx\csname citenamefont\endcsname\relax
  \def\citenamefont#1{#1}\fi
\expandafter\ifx\csname url\endcsname\relax
  \def\url#1{\texttt{#1}}\fi
\expandafter\ifx\csname urlprefix\endcsname\relax\def\urlprefix{URL }\fi
\providecommand{\bibinfo}[2]{#2}
\providecommand{\eprint}[2][]{\url{#2}}

\bibitem{Russel} J.S. Russel, in {\it 14th Mtg. of the British Assoc.
for the Advance of Science} (Murray, London, 1845), pp. 311-390, 57 plates.

\bibitem{DB_rev} S. Aubry, Physica D {\bf 103}, 201 (1997);
S. Flach, C.R. Willis, Phys. Rep. {\bf 295}, 181 (1998);
{\it Energy Localisation and Transfer}, edited by T. Dauxois,
A. Litvak-Hinenzon, R. MacKay and A. Spanoudaki
(World Scientific, Singapore, 2004);
D. K. Campbell, S. Flach and Yu. S. Kivshar,
Physics Today {\bf 57}(1), 43 (2004).

\bibitem[{\citenamefont{Eisenberg et~al.}(1998)\citenamefont{Eisenberg,
  Silberberg, Morandotti, Boyd, and Aitchison}}]{ESM+98}
\bibinfo{author}{\bibfnamefont{H.}~\bibnamefont{Eisenberg}},
  \bibinfo{author}{\bibfnamefont{Y.}~\bibnamefont{Silberberg}},
  \bibinfo{author}{\bibfnamefont{R.}~\bibnamefont{Morandotti}},
  \bibinfo{author}{\bibfnamefont{A.}~\bibnamefont{Boyd}}, \bibnamefont{and}
  \bibinfo{author}{\bibfnamefont{J.}~\bibnamefont{Aitchison}},
  \bibinfo{journal}{Phys. Rev. Lett.} \textbf{\bibinfo{volume}{81}},
  \bibinfo{pages}{3383} (\bibinfo{year}{1998}).

\bibitem[{\citenamefont{Sukhorukov et~al.}(2003)\citenamefont{Sukhorukov,
  Kivshar, Eisenberg, and Silberberg}}]{SKE+03}
\bibinfo{author}{\bibfnamefont{A.}~\bibnamefont{Sukhorukov}},
  \bibinfo{author}{\bibfnamefont{Y.}~\bibnamefont{Kivshar}},
  \bibinfo{author}{\bibfnamefont{H.}~\bibnamefont{Eisenberg}},
  \bibnamefont{and}
  \bibinfo{author}{\bibfnamefont{Y.}~\bibnamefont{Silberberg}},
  \bibinfo{journal}{IEEE J. Quantum Electron.} \textbf{\bibinfo{volume}{39}},
  \bibinfo{pages}{31} (\bibinfo{year}{2003}).

\bibitem[{\citenamefont{Fleischer et~al.}(2003)\citenamefont{Fleischer, Segev,
  Efremidis, and Christodoulides}}]{FSE+03}
\bibinfo{author}{\bibfnamefont{J.}~\bibnamefont{Fleischer}},
  \bibinfo{author}{\bibfnamefont{M.}~\bibnamefont{Segev}},
  \bibinfo{author}{\bibfnamefont{N.}~\bibnamefont{Efremidis}},
  \bibnamefont{and}
  \bibinfo{author}{\bibfnamefont{D.}~\bibnamefont{Christodoulides}},
  \bibinfo{journal}{Nature} \textbf{\bibinfo{volume}{422}},
  \bibinfo{pages}{147} (\bibinfo{year}{2003}).

\bibitem[{\citenamefont{Trias et~al.}(2000)\citenamefont{Trias, Mazo, and
  Orlando}}]{TMO00}
\bibinfo{author}{\bibfnamefont{E.}~\bibnamefont{Trias}},
  \bibinfo{author}{\bibfnamefont{J.}~\bibnamefont{Mazo}}, \bibnamefont{and}
  \bibinfo{author}{\bibfnamefont{T.}~\bibnamefont{Orlando}},
  \bibinfo{journal}{Phys. Rev. Lett.} \textbf{\bibinfo{volume}{84}},
  \bibinfo{pages}{741} (\bibinfo{year}{2000}).

\bibitem[{\citenamefont{Sato et~al.}(2003{\natexlab{a}})\citenamefont{Sato,
  Hubbard, Sievers, Ilic, Czaplewski, and Craighead}}]{SHS+03}
\bibinfo{author}{\bibfnamefont{M.}~\bibnamefont{Sato}},
  \bibinfo{author}{\bibfnamefont{B.}~\bibnamefont{Hubbard}},
  \bibinfo{author}{\bibfnamefont{A.}~\bibnamefont{Sievers}},
  \bibinfo{author}{\bibfnamefont{B.}~\bibnamefont{Ilic}},
  \bibinfo{author}{\bibfnamefont{D.}~\bibnamefont{Czaplewski}},
  \bibnamefont{and}
  \bibinfo{author}{\bibfnamefont{H.}~\bibnamefont{Craighead}},
  \bibinfo{journal}{Phys. Rev. Lett.} \textbf{\bibinfo{volume}{90}},
  \bibinfo{pages}{044102} (\bibinfo{year}{2003}{\natexlab{a}}).

\bibitem[{\citenamefont{Sato et~al.}(2003{\natexlab{b}})\citenamefont{Sato,
  Hubbard, English, Sievers, Ilic, Czaplewski, and Craighead}}]{SHE+03}
\bibinfo{author}{\bibfnamefont{M.}~\bibnamefont{Sato}},
  \bibinfo{author}{\bibfnamefont{B.}~\bibnamefont{Hubbard}},
  \bibinfo{author}{\bibfnamefont{L.}~\bibnamefont{English}},
  \bibinfo{author}{\bibfnamefont{A.}~\bibnamefont{Sievers}},
  \bibinfo{author}{\bibfnamefont{B.}~\bibnamefont{Ilic}},
  \bibinfo{author}{\bibfnamefont{D.}~\bibnamefont{Czaplewski}},
  \bibnamefont{and}
  \bibinfo{author}{\bibfnamefont{H.}~\bibnamefont{Craighead}},
  \bibinfo{journal}{Chaos} \textbf{\bibinfo{volume}{13}}, \bibinfo{pages}{702}
  (\bibinfo{year}{2003}{\natexlab{b}}).

\bibitem[{\citenamefont{Sato et~al.}(2004)\citenamefont{Sato, Hubbard, Sievers,
  Ilic, and Craighead}}]{SHS+04}
\bibinfo{author}{\bibfnamefont{M.}~\bibnamefont{Sato}},
  \bibinfo{author}{\bibfnamefont{B.}~\bibnamefont{Hubbard}},
  \bibinfo{author}{\bibfnamefont{A.}~\bibnamefont{Sievers}},
  \bibinfo{author}{\bibfnamefont{B.}~\bibnamefont{Ilic}}, \bibnamefont{and}
  \bibinfo{author}{\bibfnamefont{H.}~\bibnamefont{Craighead}},
  \bibinfo{journal}{Europhys. Lett.} \textbf{\bibinfo{volume}{66}},
  \bibinfo{pages}{318} (\bibinfo{year}{2004}).

\bibitem[{\citenamefont{Schwarz et~al.}(1999)\citenamefont{Schwarz, English,
  and Sievers}}]{SES99}
\bibinfo{author}{\bibfnamefont{U.}~\bibnamefont{Schwarz}},
  \bibinfo{author}{\bibfnamefont{L.}~\bibnamefont{English}}, \bibnamefont{and}
  \bibinfo{author}{\bibfnamefont{A.}~\bibnamefont{Sievers}},
  \bibinfo{journal}{Phys. Rev. Lett.} \textbf{\bibinfo{volume}{83}},
  \bibinfo{pages}{223} (\bibinfo{year}{1999}).

\bibitem[{\citenamefont{Sato and Sievers}(2004)}]{SS04}
\bibinfo{author}{\bibfnamefont{M.}~\bibnamefont{Sato}} \bibnamefont{and}
  \bibinfo{author}{\bibfnamefont{A.}~\bibnamefont{Sievers}},
  \bibinfo{journal}{Nature} \textbf{\bibinfo{volume}{432}},
  \bibinfo{pages}{486} (\bibinfo{year}{2004}).

\bibitem[{\citenamefont{Machida and Koyama}(2004)}]{MK04}
\bibinfo{author}{\bibfnamefont{M.}~\bibnamefont{Machida}} \bibnamefont{and}
  \bibinfo{author}{\bibfnamefont{T.}~\bibnamefont{Koyama}},
  \bibinfo{journal}{Phys. Rev. B} \textbf{\bibinfo{volume}{70}},
  \bibinfo{pages}{024523} (\bibinfo{year}{2004}).

\bibitem[{\citenamefont{Kourakis and Shukla}(2005)}]{KS05}
\bibinfo{author}{\bibfnamefont{I.}~\bibnamefont{Kourakis}} \bibnamefont{and}
  \bibinfo{author}{\bibfnamefont{P.}~\bibnamefont{Shukla}},
  \bibinfo{journal}{Phys. Plasmas}
  \textbf{\bibinfo{volume}{12}}, \bibinfo{pages}{014502}
  (\bibinfo{year}{2005}).

\bibitem[{\citenamefont{Flach}(1998)}]{Fla98}
\bibinfo{author}{\bibfnamefont{S.}~\bibnamefont{Flach}},
  \bibinfo{journal}{Phys. Rev. E} \textbf{\bibinfo{volume}{58}},
  \bibinfo{pages}{R4116} (\bibinfo{year}{1998}).

\bibitem[{\citenamefont{Rosenau and Hyman}(1993)}]{RH93}
\bibinfo{author}{\bibfnamefont{P.}~\bibnamefont{Rosenau}} \bibnamefont{and}
  \bibinfo{author}{\bibfnamefont{J.}~\bibnamefont{Hyman}},
  \bibinfo{journal}{Phys. Rev. Lett.} \textbf{\bibinfo{volume}{70}},
  \bibinfo{pages}{564} (\bibinfo{year}{1993}).

\bibitem[{\citenamefont{Rosenau}(1994)}]{Ros94}
\bibinfo{author}{\bibfnamefont{P.}~\bibnamefont{Rosenau}},
  \bibinfo{journal}{Phys. Rev. Lett.} \textbf{\bibinfo{volume}{73}},
  \bibinfo{pages}{1737} (\bibinfo{year}{1994}).

\bibitem[{\citenamefont{Kivshar}(1993)}]{Kiv93}
\bibinfo{author}{\bibfnamefont{Y.}~\bibnamefont{Kivshar}},
  \bibinfo{journal}{Phys. Rev. E} \textbf{\bibinfo{volume}{48}},
  \bibinfo{pages}{R43} (\bibinfo{year}{1993}).

\bibitem[{\citenamefont{Flach}(1994)}]{Fla94}
\bibinfo{author}{\bibfnamefont{S.}~\bibnamefont{Flach}},
  \bibinfo{journal}{Phys. Rev. E} \textbf{\bibinfo{volume}{50}},
  \bibinfo{pages}{3134} (\bibinfo{year}{1994}).

\bibitem[{\citenamefont{Dey et~al.}(2002)\citenamefont{Dey, Eleftheriou, Flach,
  and Tsironis}}]{DEF+02}
\bibinfo{author}{\bibfnamefont{B.}~\bibnamefont{Dey}},
  \bibinfo{author}{\bibfnamefont{M.}~\bibnamefont{Eleftheriou}},
  \bibinfo{author}{\bibfnamefont{S.}~\bibnamefont{Flach}}, \bibnamefont{and}
  \bibinfo{author}{\bibfnamefont{G.}~\bibnamefont{Tsironis}},
  \bibinfo{journal}{Phys. Rev. E} \textbf{\bibinfo{volume}{65}},
  \bibinfo{pages}{017601} (\bibinfo{year}{2002}).

\bibitem[{\citenamefont{Rosenau and Schochet}(2005)}]{RS05}
\bibinfo{author}{\bibfnamefont{P.}~\bibnamefont{Rosenau}} \bibnamefont{and}
  \bibinfo{author}{\bibfnamefont{S.}~\bibnamefont{Schochet}},
  \bibinfo{journal}{Phys. Rev. Lett.} \textbf{\bibinfo{volume}{94}},
  \bibinfo{pages}{045503} (\bibinfo{year}{2005});
  P.~Rosenau and S.~Schochet, Chaos {\bf 15}, 015111 (2005).

\bibitem{fw93fwo94}
S. Flach and C. R. Willis, Phys. Lett. A {\bf 181}, 232 (1993);
S. Flach, C. R. Willis and E. Olbrich,
Phys. Rev. E {\bf 49}, 836 (1994).

\bibitem{JGC+98} Yu.B.~Gaididei, S.F.~Mingaleev, P.L.~Christiansen, and
K.\O.~Rasmussen, Phys. Rev. E {\bf 55}, 6141 (1997); M.~Johansson, Yu.B.~Gaididei, 
P.L.~Christiansen, and K.\O.~Rasmussen, Phys. Rev. E {\bf 57}, 4739 (1998);
P.L.~Christiansen, Yu.B.~Gaididei, F.~Mertens, and S.F.~Mingaleev, Eur. Phys. J. B 
{\bf 19}, 545 (2001).

\bibitem{weFPU} S.~Flach, and A.~Gorbach, Chaos \textbf{15}, 015112 (2005).
  
\bibitem[{\citenamefont{Flach}(2004)}]{Fla04}
\bibinfo{author}{\bibfnamefont{S.}~\bibnamefont{Flach}}, in
  \emph{\bibinfo{booktitle}{Energy Localization and Transfer}}, edited by
  \bibinfo{editor}{\bibfnamefont{T.}~\bibnamefont{Dauxois}},
  \bibinfo{editor}{\bibfnamefont{A.}~\bibnamefont{Litvak-Hinenzon}},
  \bibinfo{editor}{\bibfnamefont{R.}~\bibnamefont{MacKay}}, \bibnamefont{and}
  \bibinfo{editor}{\bibfnamefont{A.}~\bibnamefont{Spanoudaki}}
  (\bibinfo{publisher}{World Scientific, Singapore}, \bibinfo{year}{2004}), pp.
  \bibinfo{pages}{1--71}.
  
\bibitem{stab_inv} T. Cretegny, Ph. D. thesis, \'Ecole Normale Sup\'erieure de Lyon, France 1998; 
M. \"Oster, M. Johansson, and A. Eriksson, Phys. Rev. E \textbf{67}, 056606 (2003); A.V. Gorbach and M. Johansson,  
Phys. Rev. E  \textbf{67}, 066608 (2003).

\bibitem{Chen} D.~Chen, S.~Aubry, and G.P.~Tsironis, Phys. Rev. Lett. {\bf 77}, 4776 (1996).

\end{thebibliography}

\end{document}